\documentclass[twocolumn,showpacs,prd]{revtex4}
\usepackage{graphicx}
\usepackage{color}
\usepackage{amssymb}
\usepackage{enumitem}
\usepackage[off]{auto-pst-pdf}
\definecolor{darkgreen}{rgb}{0.0,0.4,0.0}
\definecolor{paillec}{rgb}{1,1,0.8}
\definecolor{mauve}{rgb}{1,0.7,1}
\definecolor{vertbleu}{rgb}{0.4,1,0.6}
\definecolor{bleupale}{rgb}{0.4,0.8,0.8}  
\definecolor{rose}{rgb}{1,0.4,0.4} 
\definecolor{paille+}{rgb}{1,1,0.4}
\definecolor{paille++}{rgb}{1,0.8,0.4} 
\definecolor{bleuciel}{rgb}{0.4,1,1} 
\definecolor{melon}{rgb}{1,0.6,0.4} 
\definecolor{violetclair}{rgb}{0.6,0.6,1}

\newcommand{\smr}[1]{\mbox{\scriptsize #1}}

\begin{document}

\title{Strong evidence of \mbox{\boldmath $\rho(1250)$} from a unitary
multichannel reanalysis of elastic scattering data with crossing-symmetry
constraints}

\author{N. Hammoud$^a$, R. Kami\'nski$^a$, V. Nazari$^b$, G. Rupp$^c$}

\affiliation{$^a$Institute of Nuclear Physics, Polish Academy of Sciences,
Krak\'ow, Poland\\
$^b$Joint Institute for Nuclear Research, Dubna, Russia\\
$^c$Centro de F\'{\i}sica e Engenharia de Materiais
Avan\c{c}ados, Instituto Superior T\'{e}cnico, Universidade de Lisboa,
P-1049-001, Portugal}

\begin{abstract}
An analysis is presented of elastic $P$-wave $\pi\pi$ phase shifts and 
inelasticities up to 2~GeV, aimed at identifying  the corresponding
$J^{PC}=1^{--}$ excited $\rho$ resonances and focusing on the
$\rho(1250)$ vs.\ $\rho(1450)$ controversy. The approach employs an
improved parametrization in terms of a manifestly unitary and analytic
three-channel $S$-matrix with its complex-energy pole positions. The
included channels are $\pi\pi$, $\rho2\pi$, and $\rho\rho$, the latter
two being effective in the sense that they mimic several experimentally
observed decay modes with nearby thresholds. In an alternative fit, the
$\rho2\pi$ mode is replaced by $\omega\pi$, which is also an experimentally
relevant channel. The improvement with respect to prior work amounts to
the enforcement of maximum crossing symmetry through once-subtracted
dispersion relations called GKPY equations. A separate analysis concerns
the pion electromagnetic form factor, which again demonstrates the enormous
importance of guaranteeing unitarity and analyticity when dealing with
very broad and highly inelastic resonances. In the case of $\rho(1250)$
vs.\ $\rho(1450)$, the failure to do so is shown to give rise to an error
in the predicted mass of about 170~MeV. A clear picture emerges
from these analyses, identifying five vector $\rho$ states  below 2~GeV,
viz.\ $\rho(770)$, $\rho(1250)$, $\rho(1450)$, $\rho(1600)$, and
$\rho(1800)$, with $\rho(1250)$ being indisputably the most important
excited $\rho$ resonance. The stability of the fits as well as the
imposition of unitarity, analyticity, and approximate crossing symmetry
in the analyses lend very strong support to these assignments. The
possibly far-reaching consequences for meson spectroscopy are discussed.

\end{abstract}



\maketitle

\section{Introduction}
\label{intro}
The experimental status of meson resonances with masses ranging from
1 to 2 GeV is very poor. Many states expected from the quark model have
not been observed so far, whereas several apparently normal resonances
listed in the PDG tables \cite{PDG2020} do not fit in with mainstream
quark models like, for instance, the relativized meson model by Godfrey
and Isgur (GI) \cite{PRD32p189}. In Ref.~\cite{APPBPS5p1007} some of the
obvious discrepancies were briefly reviewed, e.g.\ concerning the many
observed $f_2$ states (with $J^{PC}=2^{++}$) to be contrasted with the
much fewer ones predicted in the GI model. Another disagreement is the
relatively low mass of the strange vector meson $K^\star(1410)$, the
first radial excitation of $K^\star(892)$, which is predicted almost
200 MeV higher in the GI and similar quark models. On the other hand,
$\rho^\prime$, the first radial excitation of $\rho(770)$, is listed by
the PDG as $\rho(1450)$ \cite{PDG2020}, which is difficult to reconcile
with a lighter $K^\star(1410)$, as the latter state contains one strange
quark and one light quark instead of two light quarks. However, under the
$\rho(1450)$ entry in the PDG meson listings one finds a large variety
of experimental observations, with a huge mass range of 1208--1624 MeV,
also depending on the particular strong decay mode. As a matter of
fact, there have been many indications of a lighter $\rho^\prime$,
roughly in the range 1.25--1.3 GeV, which we shall examine in more detail
in the next section.

The importance of more accurately knowing the $\rho^\prime$ mass, and
of course that of $K^{\star\prime}$, should not be underestimated. Quark
models based on the usual Coulomb-plus-linear interquark potential, with
a running strong coupling constant in the Coulombic part, predict
increasing radial splittings for lighter quarks. Therefore, a $\rho^\prime$
mass significantly lower than the value predicted in the GI and largely
equivalent models would pose a serious challenge to such approaches. Now, 
it is true that the precise mass of a broad resonance like $\rho^\prime$
depends on the way the corresponding scattering data are analyzed. In that
respect, the usual Breit-Wigner (BW) parametrizations can be very unreliable,
possibly leading to deviations of the order of 10 MeV for $\rho(770)$ and
even more than 100 MeV in the case of $\rho^\prime$ \cite{PRD96p113004}.
The reason is that multi-BW parametrizations typically do not
satisfy unitarity, which becomes a very serious issue in highly inelastic
processes like those in which $\rho^\prime$ is observed. Another problem,
this time on the theoretical side, is the usual static approach to mesonic
resonances in most quark models, treating them as manifestly stable
bound states of a quark and an antiquark. Coupled-channel effects from
meson loops, due to both open and closed meson-meson decay channels, may
give rise to sizable mass shifts, alongside producing a large width.
For instance, in the multichannel unitarized quark model of
Ref.~\cite{PRD27p1527} a bare $\rho^\prime$ mass of 1.48~GeV is
lowered by about 190~MeV owing to coupled channels, resulting in a physical
$\rho^\prime$ mass of 1.29 GeV. However, in the latter model a completely
different confinement interaction is employed, leaving doubts about the
precise size of such unitarization effects in mainstream quark models.
Moreover, also the ground states as well as the higher radial excitations
may suffer considerable mass shifts \cite{PRD27p1527}, so that a refit of
all parameters will have to be carried out when unitarizing any particular
model.

A further important piece of information comes from recent
lattice-QCD calculations. In Ref.~\cite{ARXIV13116579} masses of
light and strange hadrons were computed in an unquenched simulation,
but with only (anti)quark interpolators included and so resulting in
purely real spectra. Thus, the mass of the first radial excitation of
$K^\star(892)$ was found to be slightly above 1.6~GeV. On the other
hand, in Ref.~\cite{PRD88p054508} members of the same lattice
collaboration study $P$-wave $K\pi$ scattering in a simulation with
both quark-antiquark and meson-meson interpolators. Here, they employ
L\"{u}scher's \cite{NPB354p531} method to extract $K\pi$ phase shifts
from discrete energy levels for different lattice sizes. The resulting
phases are in good agreement with the data, including the $K^\star(892)$
mass and even its width in an extrapolation to the physical pion mass. The
big surprise is the mass of the first radial excitation of $K^\star(892)$,
coming out at $(1.33\pm0.02)$ GeV, that is, about 300 MeV lower than in
the former lattice simulation, without meson-meson interpolators.
Admittedly, the latter calculation amounts to an approximation, as only
the $K\pi$ decay channel is included, thus treating the $K^{\star\prime}$
as a purely elastic resonance. Nevertheless, the importance of accounting
for unitarity when doing spectroscopy in a quantitatively reliable
fashion is unmistakable corroborated.

The present paper aims at clarifying the status of $\rho^\prime$ and also
the higher vector $\rho$ excitations, by reanalyzing old data on 
$\pi\pi$ scattering, viz.\ elastic phase shifts and inelasticities up
to about 2 GeV. The employed model of analysis is a manifestly unitary
three-channel $S$-matrix parametrization, in which the complex pole
positions of the different $\rho$ resonances are explicitly included
through generalized BW-type expressions. Moreover, (approximate) crossing
symmetry is enforced by minimizing in the fits the difference between
the experimental real parts of amplitudes and the theoretical ones
resulting from dispersion relations. The three included channels are
$\pi\pi$, as well as the effective channels $\rho2\pi$ and
$\rho\rho$, with the latter ones mimicking $4\pi$ final states.
For further details, see Sec.~\ref{crossing}.

The paper's organization is as follows. Section~\ref{rho} extensively reviews
the status of $\rho(1250)$ vs.\ $\rho(1450)$ in decades of literature.
Section~\ref{crossing} describes the done $S$-matrix analyses of $\pi\pi$
phase shifts and inelasticities with the imposed crossing-symmetry constraints.
In Sec.~\ref{em} an analysis of the pion electromagnetic form factor further
illustrates the necessity of a unitary and analytic approach to very broad
inelastic resonances. Section~\ref{conclusions} is devoted to a general
discussion of the results and conclusions.

\section{\boldmath{$\rho(1250)$} vs.\ \boldmath{$\rho(1450)$}
in experiment and models}
\label{rho}

The first time a $\rho^\prime$ resonance was included in the PDG tables
dates back to 1974 \cite{PDG1974}, with the entry in the data-card listings
called $\rho^\prime(1250)$. Its mass and width were listed as 1256~MeV and
130~MeV, respectively, from Ref.~\cite{NPB47p61}. In the latter paper, 
evidence was found of two opposite-parity spin-1 $\omega\pi$ resonances at
about 1250~MeV in $\bar{p}p$ annihilations at rest. The vector state could
correspond to the $\rho^\prime$ and the pseudo-vector ($J^{PC}=1^{+-}$)
to what is nowadays called $b_1(1235)$ \cite{PDG2020}. To our knowledge, 
though, the earliest indication of a possible $\rho^\prime$ goes back to
1970 and was reported in Ref.~\cite{PRD1p27}, also cited in
Ref.~\cite{PDG1974}. In this experiment, neutral bosons were observed in
photoproduction on protons, including a resonance at about 1240~MeV with a
width of roughly 100~MeV \cite{PRD1p27}. The authors tentatively identified
this state with the pseudo-vector ``$B$'' (now $b_1(1235)$) meson. However, 
not having determined the $J^{PC}$ quantum numbers, they concluded: \begin{quote}
\em ``It is possible that this particle could be an as-yet-undiscovered
vector meson'' \em \/.
\end{quote}
No further vector mesons were found in the
energy region 1.3--2.0~GeV with a cross section larger than 5\% of that of
$\rho^0(770)$ (90\% confidence level \cite{PRD1p27}).

Over the following years several experiments and analyses
\cite{LNC8p659,NPB76p375,PLB52p493,NCA39p374,NCA49p207,ZPC4p169,PLB92p211,JETPL37p733}
reported a possible $\rho^\prime$ roughly in the range 1.2--1.3~GeV. The most
affirmative identification was the $\omega\pi^0$ enhancement observed
\cite{PLB92p211} in photoproduction on protons, with mass $\sim\!1.25$~GeV,
width $\sim\!0.3$~GeV, and a dominant vector component. Note that this paper
is cited in the PDG \cite{PDG2020} under the $\rho(1450)$ entry, in spite of
the 200-MeV mass discrepancy. Also referred there \cite{PDG2020} is 
Ref.~\cite{ZPC4p169}, concerning the $\omega\pi$ mode as well.

In 1982 \cite{ZPC13p43} and 1983 \cite{PRD27p1527}, 1982 two
quark-model calculations \cite{PRD27p1527,ZPC13p43}
reported support for a $\rho^\prime$ below 1.3~GeV, while also describing its
small $\pi\pi$ cross section. Starting with Ref.~\cite{PRD27p1527},
this above-mentioned unitarized multichannel quark model
\cite{PRD27p1527}  was applied to vector and pseudoscalar mesons. The resulting
$P$-wave $\pi\pi$ cross section reasonably reproduced the $\rho(770)$ mass,
width, and pole position, while also predicting a $\rho^\prime$ pole with
a real part of 1.29~GeV, though with a too small width. At this energy,
the $\pi\pi$ cross section revealed a very small enhancement on top of the
$\rho(770)$ tail (see Ref.~\cite{PRD27p1527}, FIG.~4). Furthermore, the
relativistic quark model of Ref.~\cite{ZPC13p43} predicted a $\rho^\prime$
at about 1.22~GeV, explaining its small $\pi\pi$ width as a combined
effect of the nodal structure of the radially excited $\rho^\prime$ and the
Lorentz-contracted wave functions of the outgoing two pions. Also of
interest is the amplitude analysis of the coupled $\omega\pi,\pi\pi$ system
in Ref.~\cite{ZPC29p107}, which explained the difficulty to observe a
$\rho^\prime$ in $\pi\pi$ scattering between 1.1 and 1.3~GeV as being
due to a small yet dominant inelastic background in that energy region.

The $\rho(1250)$ entry in the PDG was maintained up till 1986 \cite{PDG1986}.
Things changed in 1987 with a combined analysis \cite{ZPC33p407} of two-pion
and four-pion data from $\gamma p$ and $e^+e^-$ processes, resulting in the
postulation of two excited $\rho$ resonances, with the masses 1.465
and 1.7~GeV. Curiously, no $\rho^\prime$
below 1.3~GeV was even considered in the different fits to the data, in
spite of its systematic inclusion in the PDG tables since 1974.
As a matter of fact, no mention at all of such a resonance was made
in Ref.~\cite{ZPC33p407}. On the other hand, the GI relativized quark
model \cite{PRD32p189} mentioned above was explicitly cited, for having
predicted the $\rho^\prime$ at a mass of 1.45~GeV and the corresponding
$1\,^{3\!}D_1$ state at 1.66~GeV, i.e., values very close to the ones found
in these fits \cite{ZPC33p407}. A key feature in the latter analysis is the
authors' (unproven) conjecture, in the context of the vector-meson-dominance
model for photoproduction, that the off-diagonal transitions
$1\,^{3\!}S_1 \to 2\,^{3\!}S_1$ and $1\,^{3\!}S_1 \to 1\,^{3\!}D_1$ 
have cross sections comparable to the transitions
$1\,^{3\!}S_1 \to 1\,^{1\!}P_1$ and $1\,^{3\!}S_1 \to 1\,^{3\!}D_3$, which
correspond to the diffractive photoproduction of the $1^{+-}$
``$B(1250)$'' meson (now called $b_1(1235)$) and $3^{--}$ ``g(1690)''
(now $\rho_3(1690)$ \cite{PDG2020}), respectively. The subsequent 1988
PDG edition \cite{PDG1988} then included the new entries $\rho(1450)$
and $\rho(1700)$, while completely eliminating the $\rho(1250)$ and
oddly accommodating the $\rho(1250)$ observation of Ref.~\cite{PLB92p211}
under $\rho(1450)$. This state of affairs has remained unaltered so far
\cite{PDG2020}.

One of the authors of Ref.~\cite{ZPC33p407} published
several more papers on the $\rho^\prime$ and related issues, which merit 
some further attention. The first of these appeared in 1991
\cite{PLB269p450}, in a reaction to a new observation \cite{NPBPS21p105}
of a $\rho^\prime$ below 1.3~GeV, with mass 1266~MeV and width 166~MeV.
These vector-resonance parameters resulted from a partial-wave amplitude
analysis of the $\pi^+\pi^-$ system observed with the LASS spectrometer
at SLAC for the reaction $K^-p\to\pi^+\pi^-\Lambda$ at 11 GeV.
In Ref.~\cite{PLB269p450} the authors claimed that
\begin{quote}
\em
``the interpretation of the LASS state at 1.27~GeV as a radial excitation
of the $\rho$ can in all probability be ruled out on the basis of its 
very small electromagnetic coupling'' \em. 
\end{quote}
Note, however, that
Ref.~\cite{NPBPS21p105} did not make any statement about this coupling or
the associated $\rho(1270)\to e^+e^-$ width. Instead, the authors of
Ref.~\cite{PLB269p450} carried out a BW fit including 
$\rho(770)$, a ``$\rho_x$'' with fixed $m_{\rho_x}=1.27$~GeV and
$\Gamma_{\rho_x}=0.17$~GeV, a ``$\rho_1$'' with fixed
$m_{\rho_1}=1.44$~GeV and $\Gamma_{\rho_1}=0.36$~GeV, and a ``$\rho_2$''
with mass and width to be adjusted to the data, resulting in
$m_{\rho_2}=(1.73\pm0.02)$~GeV and $\Gamma_{\rho_2}=(0.29\pm0.07)$~GeV.
The problem is that such a BW description of overlapping resonances
violates $S$-matrix unitarity \cite{NCA107p2511}, which is all the more
serious for extremely broad states, as in the present case. Moreover,
the conclusion \cite{PLB269p450} that the found $\rho_x\to e^+e^-$ width
is too small was based on a comparison with GI model predictions
\cite{PRD32p189}, the very same ones that were claimed to be incompatible
with a prior \cite{ZPC33p407} analysis of $\rho^\prime$ and
$\rho^{\prime\prime}$. Another curiosity in Ref.~\cite{PLB269p450}
is the following conclusion: \em ``Thus there does not seem to be
a strong case for the interpretation of the LASS state as an exotic'' \em.
This contrasts with several later collaborative works
\cite{ZPC59p621,ZPC60p187,ZPC62p455,PRD60p114011} of one of the authors
of Ref.~\cite{PLB269p450}, which advocated the
interpretation of the LASS ``$\rho_x$'' resonance as a crypto-exotic
four-quark state (having non-exotic quantum numbers).
However, such an assignment would require \cite{ZPC60p187} the
existence of a narrow isoscalar partner state ``$\omega_x$'' at about
1.1~GeV, for which there is no experimental evidence. The very
broad $h_1(1170)$ \cite{PDG2020} is the only isoscalar $J\!=\!1$ resonance
with a nearby mass and it is easily interpreted as a normal unflavored
$1^{+-}$ $q\bar{q}$ state. 

Despite the dominant consensus on $\rho^\prime$ at 1.45~GeV and the
corresponding $1\,^{3\!}D_1$ state at 1.7~GeV, several later experiments
and analyses contradict this picture. The OBELIX \cite{PLB414p220}
spin-parity analysis of $\bar{p}p\to2\pi^+2\pi^-$ annihilations at very
low momentum resulted in the clear identification of a $\rho^\prime$
resonance with mass $(1.282\pm0.037)$~GeV and width $(0.236\pm0.036)$~GeV,
i.e., values compatible with those \cite{NPBPS21p105} of the LASS
collaboration. More recently, a combined 2-channel $S$-matrix and
generalized BW analysis of $P$-wave $\pi\pi$ phase shifts and
inelasticities up to 1.9~GeV was carried out \cite{NPA807p145}, satisfying
analyticity and multichannel unitarity. As a result, not only was a
$\rho(1250)$ firmly established for both methods of analysis, but
evidence of higher $\rho$ excitations was also found, viz.\ at roughly
1.6 and 1.9~GeV. Remarkably, a further state at about 1.45--1.47~GeV
could be accommodated as well, though its inclusion in the fits turned
out to be almost immaterial. A generalization of the mentioned
2-channel $S$-matrix parametrization to three channels in
Ref.~\cite{PRD81p016001} largely confirmed these results, the most 
significant difference being the prediction of a higher $\rho$ excitation
at about 1.8~GeV instead of 1.9~GeV. Let us finish this discussion of the
literature on $\rho(1250)$ vs.\ $\rho(1450)$ by quoting D.~V.~Bugg
\cite{PR397p257}:
\begin{quote}
\em ``It is not clear how to assign $1^-$ states to
these \em \/[Regge] \em trajectories. The $\rho(1450)$ does not seem to 
fit naturally as the first radial excitation of the $\rho(770)$.'' \em
\end{quote}

In the next section we shall outline in detail our present method of
analysis, which amounts to a further improvement of the analyses
performed in Refs.~\cite{NPA807p145} and \cite{PRD81p016001}, by 
also imposing constraints from crossing symmetry.

\section{\boldmath{$\rho(1250)$} from an analysis with crossing-symmetry
constraints}
\label{crossing}

The most recent confirmation of $\rho(1250)$, fully supported by
physical and mathematical arguments, resulted from a dispersive analysis
\cite{PRD94p116013} of the amplitudes in three coupled $P$-wave decay
channels, with a built-in crossing-symmetry condition in the
$\pi\pi$ channel. A more detailed account of this work appeared in the
PhD thesis of one of the present authors (VN) \cite{Nazari-thesis},
from which we have selected the most important results
and figures for the present paper and section.

Besides the $\pi\pi$ channel, for which experimental phase shifts and
inelasticities were available, two additional, effective channels were
included in the analysis, which should simulate the dominant three- and
four-body decays of the $\rho$ excitations. The problem here is that higher
$\rho$ resonances have many observed decay channels, which would be unfeasible
to account for completely in our $S$-matrix approach, as it would lead to a 
proliferation of Riemann sheets and complex poles. Therefore, we are guided
by the decay modes in the PDG listings \cite{PDG2020}, considerations from
channel couplings, and the $\pi\pi$ phase shifts themselves. Thus, under the
PDG $\rho(1450)$ entry, we notice the modes $a_1(1260)\pi$, $h_1(1170)\pi$,
$\pi(1300)\pi$, and $\rho(\pi\pi)_{\smr{$S$-wave}}$, all included in the $4\pi$
decays. Now, $a_1(1260)$, $h_1(1170)$, and $\pi(1300)$ all decay mostly to
$\rho\pi$, so that we have three decay channels of an excited vector $\rho$
state that do not lie very far apart and all lead to a quasi-final state of
$\rho\pi\pi$, with $4\pi$ being of course the true final state. And then there
is the $\rho(\pi\pi)_{\smr{$S$-wave}}$ channel, which will naturally be
dominated by the $f_0(500)$ (alias $\sigma$) resonance, with central threshold
in the same energy ballpark. So we mimic these decays by including one
$\rho2\pi$ channel, with threshold at 1055~MeV. Of course, the mentioned four
channels have central thresholds that lie 195--385~MeV \cite{PDG2020}
higher, but the involved resonances are extremely broad, so that opening our
effective channel at the $\rho2\pi$ threshold seems reasonable. As for the
third effective channel to be included in our analysis, we note that
an isovector vector state couples very strongly to a $P$-wave $\rho\rho$
state, which is just a matter of recoupling coefficients of spin, isospin,
and orbital angular momentum  \cite{PRD27p1527}. Indeed, under the PDG
\cite{PDG2020} $\rho(1700)$ entry, we see the $\rho\rho$
decay mode among the \em `dominant' \em\/ $\rho\pi\pi$ decays,
besides the already considered $a_1(1260)\pi$, $h_1(1170)\pi$, and
$\pi(1300)\pi$ modes. Now, the central PDG threshold mass of the $\rho\rho$
channel is 1550~MeV, but also the $\rho$ is a relatively broad resonance,
so that it is reasonable to open this channel at a somewhat lower energy.
Thus, a value of 1512~MeV was obtained already in Ref.~\cite{PRD81p016001},
upon fitting the $P$-wave $\pi\pi$ phase shifts, which is the threshold value
we keep in the present analysis, too. Note that in
Refs.~\cite{PRD81p016001,Nazari-thesis} this effective channel was called
$\rho\sigma$, but we now prefer to call it $\rho\rho$. Clearly, the
$\rho f_0(500)$ channel will also contribute to the $\rho(1700)$ decays just
as to those of $\rho(1450)$, in spite of this mode not being listed under
the $\rho(1700)$ PDG entry \cite{PDG2020}. Nevertheless, the designation
$\rho\rho$ appears more appropriate for our third effective decay channel.

These three channels were included in the analyses in
Refs.~\cite{PRD81p016001,PRD94p116013,Nazari-thesis}, the methods and results
of which we discuss below. However, when considering $\rho(1250)$ decays,
special attention should be paid to the $\omega\pi$ decay mode observed in
several experiments \cite{PDG2020}. So in order to have a more complete
analysis, we now also do a fit replacing the $\rho 2\pi$ channel by
$\omega\pi$, with threshold at 922~MeV. 
The corresponding results we will discuss below, right after presentation of methots used in our fits. 
%presenting the results of the main analysis, with the $\pi\pi$, $\rho2\pi$, and $\rho\rho$ channels.

The simple phenomenological approach in this description is not based on some
model with a phenomenological potential, but rather on a careful analysis of
the $S$-matrix poles of certain resonances on different
Riemann sheets of the complex energy plane. Among the free
parameters in the amplitudes, fitted to the experimental data and to dispersion
relations called GKPY equations \cite{PRD83p074004}, are the
complex pole positions themselves, making the obtained results largely
model-independent. This also allows to avoid problems from searching for poles
by analytically continuing the amplitudes into the complex energy or momentum
plane, as the pole postions are determined directly in the fits. 

The amplitudes are fully unitary and analytical, viz.\ of the form
\begin{equation}
\label{Eq_A}
A_{kl}(s) = \frac{1}{2i}\frac{S_{kl}-\delta_{kl}}{1-\frac{4m^2}{s}} ,
\end{equation}
where $s$ is the effective two-pion mass squared and the $S_{kl}$ are
$S$-matrix elements. 
For example, in the case of the $\pi\pi$ channel, such an element reads 
\begin{equation}
S_{11} = \eta_{11} e^{2i\delta_{11}} = S_{11}^{\smr{res}} S_1^{\smr{bgr}} = 
\frac{d_{\smr{res}}^{*}(-w^{*})} {d_{\smr{res}}(w)}
{D_{\smr{bgr}}(k_{1})},
\label{Eq_Sfull}
\end{equation}
and expressions for other matrix elements are given in Eq.~6 of
Ref.~\cite{PRD81p016001}.
Phase shifts and inelasticities are denoted by $\delta_{11}$ and $\eta_{11}$. 
For simplicity, they will be just called $\delta$ and $\eta$ further on in the text.
The $S$-matrix factors $S^{\smr{res}}$ and $S^{\smr{bgr}}$ stand for resonant
and background parts, respectively, while $d_{\smr{res}}$ are the Jost functions, which contain all
the dynamics of the interacting particles, both in individual
channels and between them. 
The momenta in a given channel are denoted by
$k_i$ and the uniformizing variable $w$ is defined as
\begin{equation}
w = \frac{\sqrt{s-s_{2}}+\sqrt{s-s_{3}}}{\sqrt{s_{3}-s_{2}}},
\label{Eq_w}
\end{equation}
where $s_2$ and $s_3$ are the thresholds of the second and third channel,
respectively. The variable $w$ transforms the eight-sheeted Riemann surface
into a simpler complex plane.
A resonance pole is given by $\sqrt{s_r}=E_r-i\Gamma_r/2$,
with $E_r$ the resonance mass and $\Gamma_r$ its full width. So for $s=s_r$,
we have
\begin{equation}
w_{r}= \frac{\sqrt{s_{r}-s_{2}}+\sqrt{s_{r}-s_{3}}}{\sqrt{s_{3}-s_{2}}}
\label{Eq_wr}
\end{equation}
and the resonance contributions $S^{\smr{res}}$ are defined as
\begin{equation}
d_{\smr{res}}(w)= w^{\frac{-M}{2}} \prod_{r=1}^{M}(w + w^{*}_{r})
\label{Eq_d}
\end{equation}
where $M$ is the number of all poles. 
The background function has the form
%\cite{PRD81p016001}
\begin{equation}
%D_{\smr{bgr}}(k_{1}) = \exp\left[-ia\,\mbox{sgn}(k_{1}) +
%b\Bigg(\frac{k_{1}}{m_{1}}\Bigg)^{3}\right], 
D_{\smr{bgr}}(k_{1})=\exp\left[2ia-2b\left(\frac{k_{1}}{m_{1}}\right)^{\!\!3}
\Theta(s,s_2)\right],
\label{Eq_bgr}
\end{equation}
with $\Theta(s,s_2)$ the Heaviside function ($=1$ for $s>s_2$) and 
where $a$ and $b$ are real numbers.

In the analysis of Ref.~\cite{PRD81p016001}, (prior to
Ref.~\cite{PRD94p116013}), the whole mathematical formalism described above
was presented, and poles connected to a given resonance yet lying on different
Riemann sheets were grouped into so-called clusters. 
These Riemann sheets depend on the number and type of analyzed
channels. In the $N$-channel case, the $S$-matrix is defined on a $2^N$-sheeted
Riemann surface. Riemann sheets are numbered according to the signs of the
imaginary parts of the relative momenta in all channels. So for three channels
there are eight Riemann sheets on which the poles can lie, and they are numbered
as follows: $I (+, +, +)$, $II (-, +, +)$, $III (-,-,+)$, $IV (+,-,+)$, 
$V (+,-,-)$,  $VI (-,-,-)$, $VII (-,+,-)$, and  $VIII (+,+,-)$.
Grouping into clusters is achieved by a weak albeit significant restriction on
the freedom of the poles' movement, always occurring as a result of the
coupling between channels. 
As a result of fits to data for phase shifts and inelasticities in the
$\pi\pi$ channel below 2~GeV (see Chapter III and FIG.~2 in
Ref.~\cite{PRD81p016001}), five resonances lying on various Riemann sheets
were thus found, namely $\rho(770)$, $\rho(1250)$, $\rho(1450)$, $\rho(1600)$,
and $\rho(1800)$. 

The same amplitudes were then used in the later works
\cite{PRD94p116013,Nazari-thesis}. However, three important changes were
introduced: 
\begin{enumerate}[label=\alph*)]
\item
the restrictions on the movement of poles as a function of coupling between
channels were removed; 
\item
the tested amplitudes (in $S$, $P$, $D$, and $F$ waves) were subjected to
the limitation resulting from fulfilling crossing symmetry in the $\pi\pi$
channel up to about 1100~MeV;
 \item
a threshold expansion with four parameters was used below about 640 MeV. 
\end{enumerate}

Removing the restrictions in point a) only lead to significant shifts of pole
positions for two of them, associated with the $\rho(770)$ cluster.
They shift by several hundred MeV and thus produce very small phase shifts
and inelasticity, typical for a weak background. It can be said that these 
become ``background poles''. Two other poles of $\rho(770)$ and all the
others of the higher $\rho$ states shift by only a few MeV or less.
The limitation in point b) is done by introducing in $\chi^2_{\smr{total}}$
a component $\chi^2_{\smr{CS}}$ corresponding to the mentioned crossing
symmetry (CS). The imposition of this symmetry is controlled by 
\begin{equation}
\label{Chi2_CS}
\chi^2_{\smr{CS}}=\sum_{i}^N \frac{[\mbox{Re}A^{\smr{(in)}}(E_i)-
                                    \mbox{Re}A^{\smr{(out)}}(E_i)]^2}{\epsilon},
\end{equation}
where $N=26$ is number of chosen energies (between the $\pi\pi$ threshold and
1100 MeV) at which $\chi^2_{\smr{CS}}$ is calculated, and $\epsilon$ is fixed
at 0.01 in order to make the size of $\chi^2_{\smr{CS}}$ comparable to the
other contributions to $\chi^2_{\smr{total}}$ (i.e., the $\chi^2_{\smr{data}}$
parts). Furthermore, $\mbox{Re}A^{\smr{(in)}}(E)$ is the real part of the
amplitude used to fit the data and the GKPY equations, while
$\mbox{Re}A^{\smr{(out)}}(E)$ is the same quantity yet calculated through the
dispersion relations
\begin{eqnarray}
\label{Eq_GKPY}
{\mbox{Re}A_{\ell}^{I{\smr{(out)}}}(s)} & = & \sum
\limits_{I'=0 }^2C_{st}^{II'}{a_0^{I'}} + \displaystyle \sum\limits_{I'=0}^2
\displaystyle \sum\limits_{\ell'=0}^4 \nonumber \\ 
& - & \hspace{-.62cm}\displaystyle\int\limits_{4m_{\pi}^2}^{\infty}\!ds'\,
K_{\ell\ell^\prime}^{I I^\prime}(s,s')\,
\mbox{Im}A_{\ell'}^{I^{\prime}\smr{(in)}}(s').
\end{eqnarray}
Here, $C_{st}^{II'}$ is the crossing matrix between $\pi\pi$ channels
with isospin $I$ and $I'$, $a_0^{I'}$ is the $S$-wave scattering-length vector
for isospin $I'$, and $K_{\ell \ell^\prime}^{I I^\prime}(s,s')$ are the
corresponding kernels for once-subtracted dispersion relations with imposed
crossing symmetry. As demonstrated in Ref.~\cite{PRD83p074004}, these
dispersion relations produce significantly smaller errors in the computed
amplitudes (actually in their real parts, from Eq.~(\ref{Eq_GKPY})) than the
well-known Roy \cite{PLB36p353} dispersion relations with two subtractions.
In practice, the integrals in Eq.~(\ref{Eq_GKPY}) are done from the $\pi\pi$
threshold up to about 2~GeV, because data are lacking at higher energies and
so-called driving terms are used thereabove.  These driving terms have the
same structure as the kernel, but are not related to the experimental input
amplitudes $A_{\ell'}^{I^{\prime{\smr{(in)}}}}(s')$. Their $s$ and $t$
dependence is given by Regge parametrization.

The amplitudes in Ref.~\cite{PRD81p016001} had a bad threshold behavior, i.e.,
they produced incorrect $\pi\pi$ scattering lengths. Nevertheless, this did
not prevent obtaining very reasonable results for the resonance pole positions
of the different $\rho$s for lying far above the $\pi\pi$ threshold.
However, when carrying out fits to the GKPY equations in
Refs.~\cite{PRD94p116013,Nazari-thesis}, the threshold behavior of the 
amplitudes became very important, since the integrals in Eq.~(\ref{Eq_GKPY})
start at threshold and the amplitudes there are the least suppressed by the
dependence of the kernels on $s$ (for explicit formulae of the kernels, see
the Appendix of Ref.~\cite{PRD83p074004}).

In order to improve the near-threshold behavior of the amplitudes (point c)
above) from Ref.~\cite{PRD81p016001}, the original amplitude in
Eqs.~(\ref{Eq_A})--(\ref{Eq_bgr}) is replaced by a polynomial below about
640~MeV (this value resulted from fits to the data and the GKPY equations).
The polynomial is merely a generalized near-threshold expansion in
powers of the pion momentum $k$, viz.\
\begin{eqnarray}
\mbox{Re}A(s) & = & \frac{\sqrt{s}}{4k}\sin 2\delta = \nonumber \\
& & m_{\pi}k^2\left[ a + bk^{2} + ck^{4} + dk^{6} + O(k^{8})\right],
\label{Eq_amp6}
\end{eqnarray}
where $a$ and $b$ are just scattering length
and effective range, respectively, which can be fixed or fitted to the data and
to the dispersion relations. The parameters $c$ and $d$ are free in the fits
to the data and to the GKPY equations, being used to smoothly match the phase
shifts from the polynomial (i.e., their values and first derivatives) to the
multichannel ones determined by Eqs.~(\ref{Eq_A})--(\ref{Eq_bgr}) at the
matching energy of about 640~MeV.
This value is still below the pole mass of $\rho(770)$, so
the effective-range approximation can be used within these limits, as opposed
to, for example, the $S$~wave and the low-lying $f_0(500)$, in which case the
matching energy must be well below 500 MeV.

This way the $\pi\pi$ $P$~wave is fitted simultaneously to the dispersion relations
and the experimental data.
The $\chi^2_{\smr{total}}$ is defined as
\begin{equation}
\label{Chi2_def}
\chi^2_{\smr{total}} = \chi^2_{\smr{CS}} + \chi^2_{\smr{Data}}\,,
\end{equation}
where $\chi^2_{\smr{CS}}$ defined in Eq.~(\ref{Chi2_CS}) includes input from
all 6 important partial waves ($JI$: $S0$, $S2$, $P1$, $D0$, $D2$, and $F1$)
and $\chi^2_{\smr{Data}}$ contains only phase shifts and inelasticities of
the $P1$ wave (hereafter just called $P$ wave). Merely this wave is adjusted
during the fits and changes to all ${\mbox{Re}A_{\ell}^{I{\smr{(out)}}}(s)}$
in Eq.~(\ref{Eq_GKPY}) are caused exclusively by modifications of this single
wave. The free parameters are: mass and width of all $P$-wave resonances, i.e.,
for $\rho(770)$ (8 parameters), $\rho(1250)$ (16 parameters), $\rho(1450)$
(8 parameters), $\rho(1600)$ (8 parameters), and $\rho(1800)$ (8 parameters),
the matching energy, and the effective-range parameter $a$ from the background
part in Eq.~(\ref{Eq_bgr}) ($b$ is fixed at $-0.85\times 10^{-4}$ to avoid
violating unitarity in the inelasticities above 1.7~GeV). Thus, the maximum
number of free parameters in the fits is 50. The parameters $a$ and $b$ in
the polynomial defined in Eq.~(\ref{Eq_amp6}) are kept fixed at the values
0.0381~$m_{\pi}^{-3}$ and 0.00523~$m_{\pi}^{-5}$, respectively.  

To avoid ending up in some local minima instead of the global one, the fits
are performed sequentially with an increasing number of free parameters,
viz.\ as follows:
\begin{itemize}
\item 1st step: only matching energy, background parameter $a$, and
$\rho(770)$, i.e., 10 free parameters;
\item 2nd step: as in the 1st step, plus $\rho(1250)$, i.e., 26 free
parameters;
\item 3rd step: as in the 2nd step, plus $\rho(1450)$, i.e., 34 free
parameters;
\item 4th step: as in the 3rd step, plus $\rho(1600)$, i.e., 42 free
parameters;
\item 5th step: as in the 4th step, plus $\rho(1800)$, i.e., 50 parameters.
\end{itemize}
Additionally, the fits are at each stage carried out repeatedly with a
different number and values of the added resonance's initial parameters.
The fitted parameters in each step serve as starting parameters in the next
step. In the first step, also the matching energy and background parameter
$a$ are taken at different initial values. The total number of employed
parameters (50) and their respective numbers for each resonance are the
smallest ones that lead to good values of $\chi^2_{\smr{total}}$. 
Increasing them further does no longer give rise to a significant
improvement in $\chi^2_{\smr{total}}$.

As already mentioned above, a relevant decay mode not
considered in our analysis so far is $\omega\pi$, which is included under both
the $\rho(1450)$ and $\rho(1700)$ PDG \cite{PDG2020} entries, with the
corresponding resonance mass ranges of 1250--1624~MeV and 1550--1800~MeV,
respectively. Our choice to do the main analysis with an effective $\rho2\pi$
channel instead of the $\omega\pi$ channel was based on the consideration that
the effective one should be more important, as it is believed to account for
several observed decay modes. Redoing our analysis while replacing the
$\rho2\pi$ channel by $\omega\pi$ just amounts to changing the threshold value
to 922~MeV, down from 1055~MeV, whereafter again fits to the experimental data
and the GKPY equations are carried out. As a result, the masses of all
resonances (i.e., the real parts of the poles) change only slightly:
$\rho(770)$ by +0.1~MeV, $\rho(1250)$ by +9.7~MeV, $\rho(1450)$ by +6.5~MeV,
$\rho(1600)$ by +2.4~MeV, and $\rho(1800)$ by -7.5~MeV.

These small changes again show the stability and reliability of the obtained
results. Moreover, comparing the quality of the two fits (values of $\chi^2$)
as presented in Table~\ref{Tab_Chi2}, we see that they are
essentially equivalent. 
The number of degrees of freedom (n.d.f.) in the fit equals 297, that is, 191
data points + 6$\times$26 (26 energies for 6 partial waves) minus 50 free
parameters.  In the fit with the $\omega\pi$ channel, the energy dependence of
the phase shift and inelasticity does not undergo any significant qualitative
modification and only small quantitative changes. 

\begin{table}[!htbp]
\centering
\begin{tabular}{|l|c|c|c|c|} 
\hline
channel & $\chi^2_{total}$ & $\chi^2_{total}/n.d.f$ & $\chi^2_{Data}$ & $\chi^2_{CS}$ \\
\hline
$\rho 2\pi$ & 403.1 & 1.357 & 294.1 & 109.0 \\
\hline
$\omega\pi$ & 406.4 & 1.368 & 296.0 & 110.4 \\
\hline
\end{tabular}
\caption{Values of $\chi^2$ for the fits with different second channel.}
\label{Tab_Chi2}
\end{table}

The following description focuses on our analysis with
$\rho 2\pi$ as the second channel.

The introduction of the three modifications a), b), and c) does not cause as
significant alterations in the $P$-wave as in the $S$-wave treated in
Ref.~\cite{PRD94p116013}, but it brings about several numerical changes. 
First of all, it is guaranteed that this new amplitude is not only unitary
and analytic like in Ref.~\cite{PRD81p016001}, but it also has the correct
threshold behavior and fulfills the crossing-symmetry condition from the
$\pi\pi$ threshold up to about 1100~MeV.
This modified amplitude, when fitted to the data given in
Ref.~\cite{PRD94p116013} as well as to the GKPY equations, results in a 
set of pole clusters from which these resonance poles, i.e.,
 the ones having by far the largest
effect on the whole amplitude, are presented in Table~\ref{Tab_poles_BKN}.
%(for the numbering convention of the 8 Riemann sheets, see e.g.\
%Refs.~\cite{PRD81p016001,Nazari-thesis}).
\begin{table}[h!]
\begin{center}
\begin{tabular}{|c|c|c|}
\hline
Resonance & Riemann Sheet & $E_{r}$ , $\Gamma_{r}/2$ (MeV) \\
\hline
& & \\
$\rho(770)$ &II & $765.2 \pm 0.4$ , $73.1 \pm 0.3$\\
& &\\
$\rho(1250)$ &III &  $1264.1 \pm 33$ ,  $146.7\pm 12$\\
& &\\
$\rho(1450)$ &III & $1424.7 \pm 26$ , $104.9 \pm 24$ \\
& & \\
$\rho(1600)$&IV & $ 1595.1 \pm 5$ , $69.5 \pm 4$ \\
& & \\
$\rho(1800)$&VI & $1779.2 \pm 14$ , $121.9 \pm 16$\\
& & \\
\hline
\end{tabular}
\end{center}
\caption{Pole positions on various Riemann sheets, for
$\sqrt{s_{r}}= E_{r}- i\Gamma_{r}/2$, of the unitary amplitude fitted
to experimental data and GKPY equations.} 
\label{Tab_poles_BKN}
\end{table}

In order to efficiently take into account the influence of all higher
$\rho$ decay channels not included in the $d_{\smr{res}}(w)$ Jost function
in Eq.~(\ref{Eq_d}), the simplest possible background is
introduced (see Eq.~(\ref{Eq_bgr})) and fitted to the data as well as the
GKPY equations. As a result, a constant and small phase of almost
-20$^\circ$ and a smoothly increasing small inelasticity are obtained.

\begin{figure}[!h]
\begin{center}
\includegraphics[angle=0,scale=0.34]{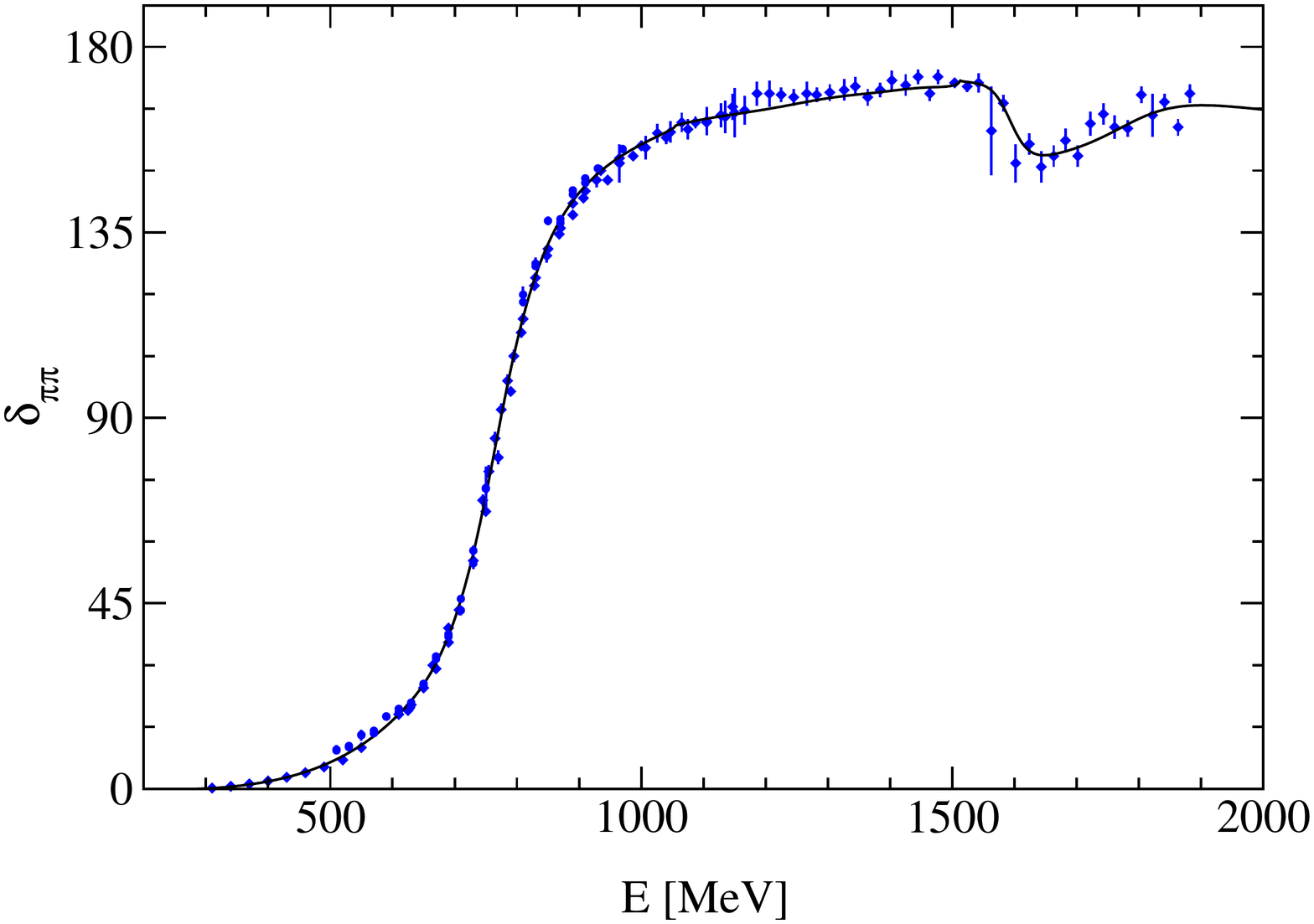}\\
\includegraphics[angle=0,scale=0.34]{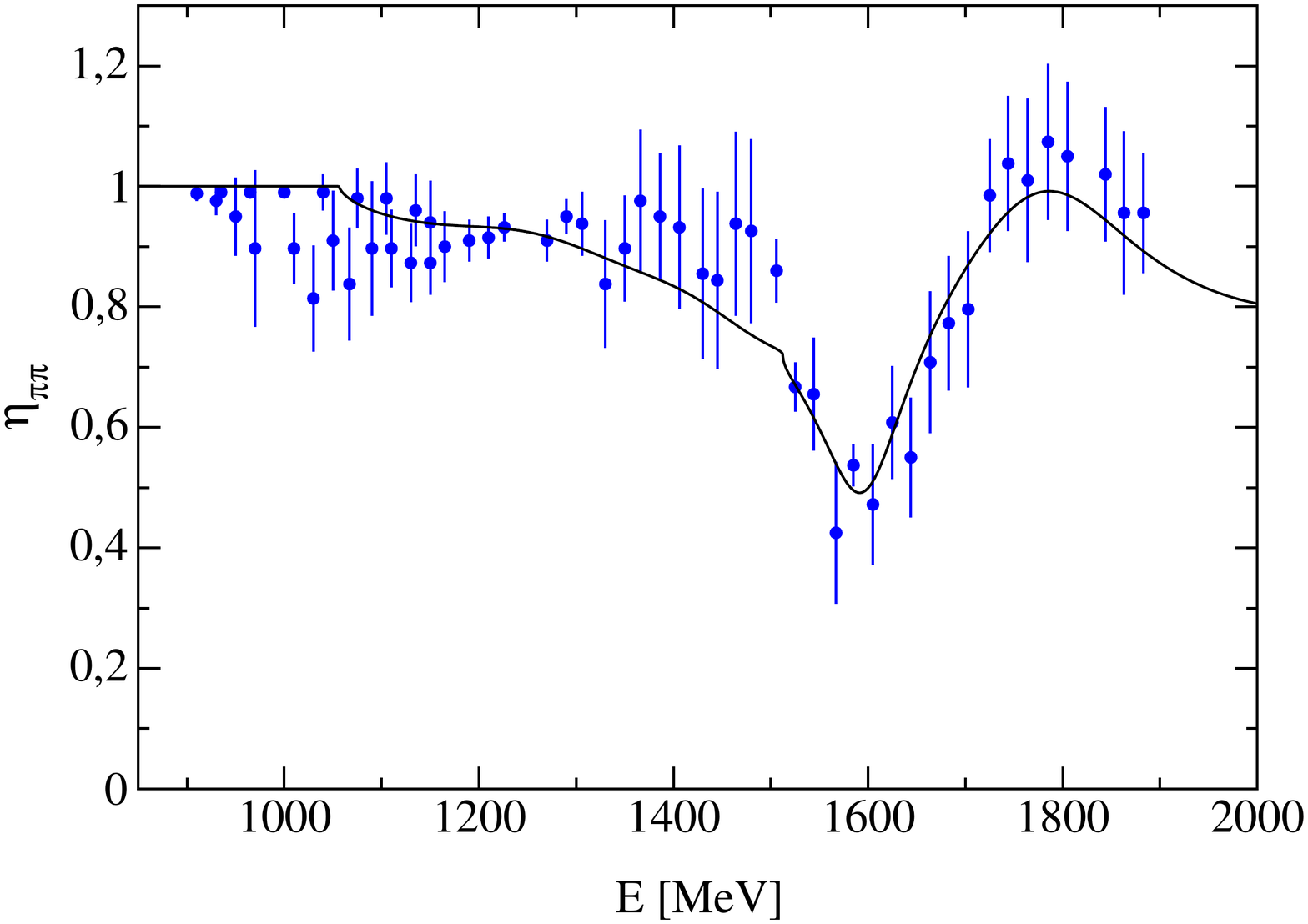}
\caption{Phase shifts and inelasticities fitted to data taken from \cite{Batley2008}-\cite{Hyams75} for the phase shifts and from \cite{Batley2008}-\cite{Hyams73},\cite{Hyams75} for inelasticity.}
\label{FitDeltaAndEtaWithData}
\end{center}
\end{figure}

\begin{figure}[!h]
\begin{center}
\includegraphics[angle=0,scale=0.34]{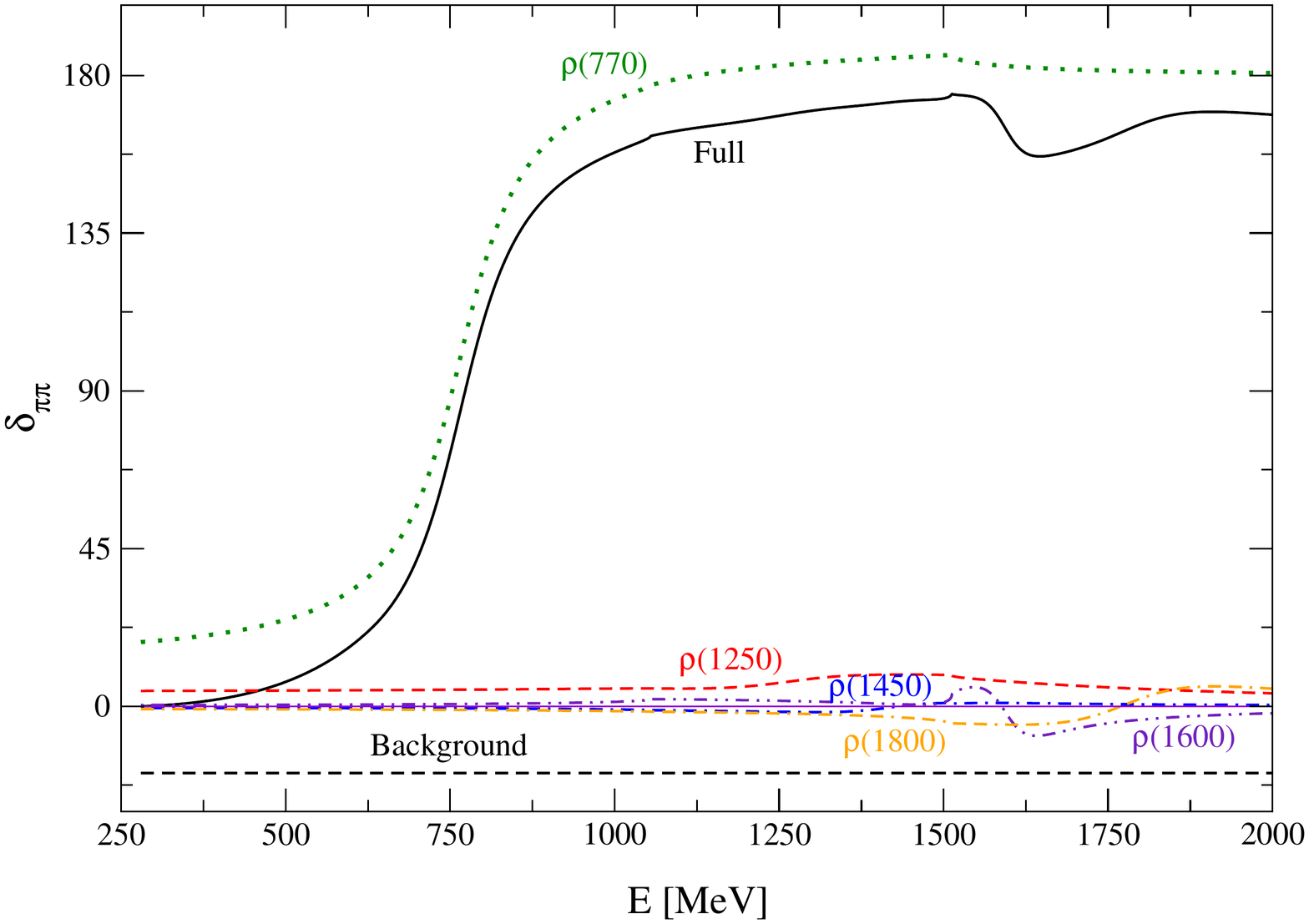}\\
\includegraphics[angle=0,scale=0.34]{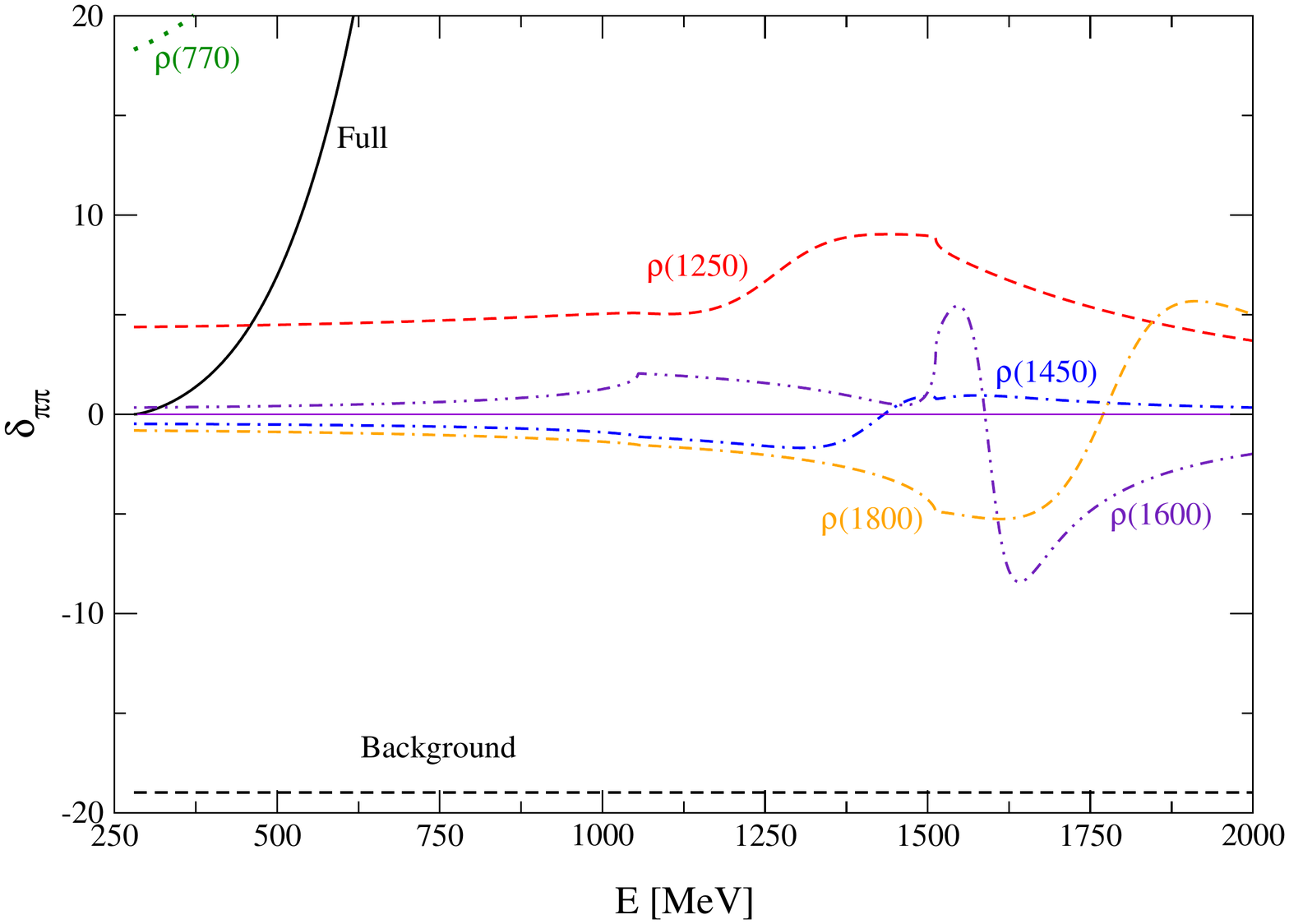}
\caption{Upper figure: phase shift due to individual resonances, as well
as full phase and background; lower figure: enlarged fragment
from the upper figure.}
\label{ClastersDelta}
\end{center}
\end{figure}

\begin{figure}[!h]
\begin{center}
\includegraphics[angle=0,scale=0.34]{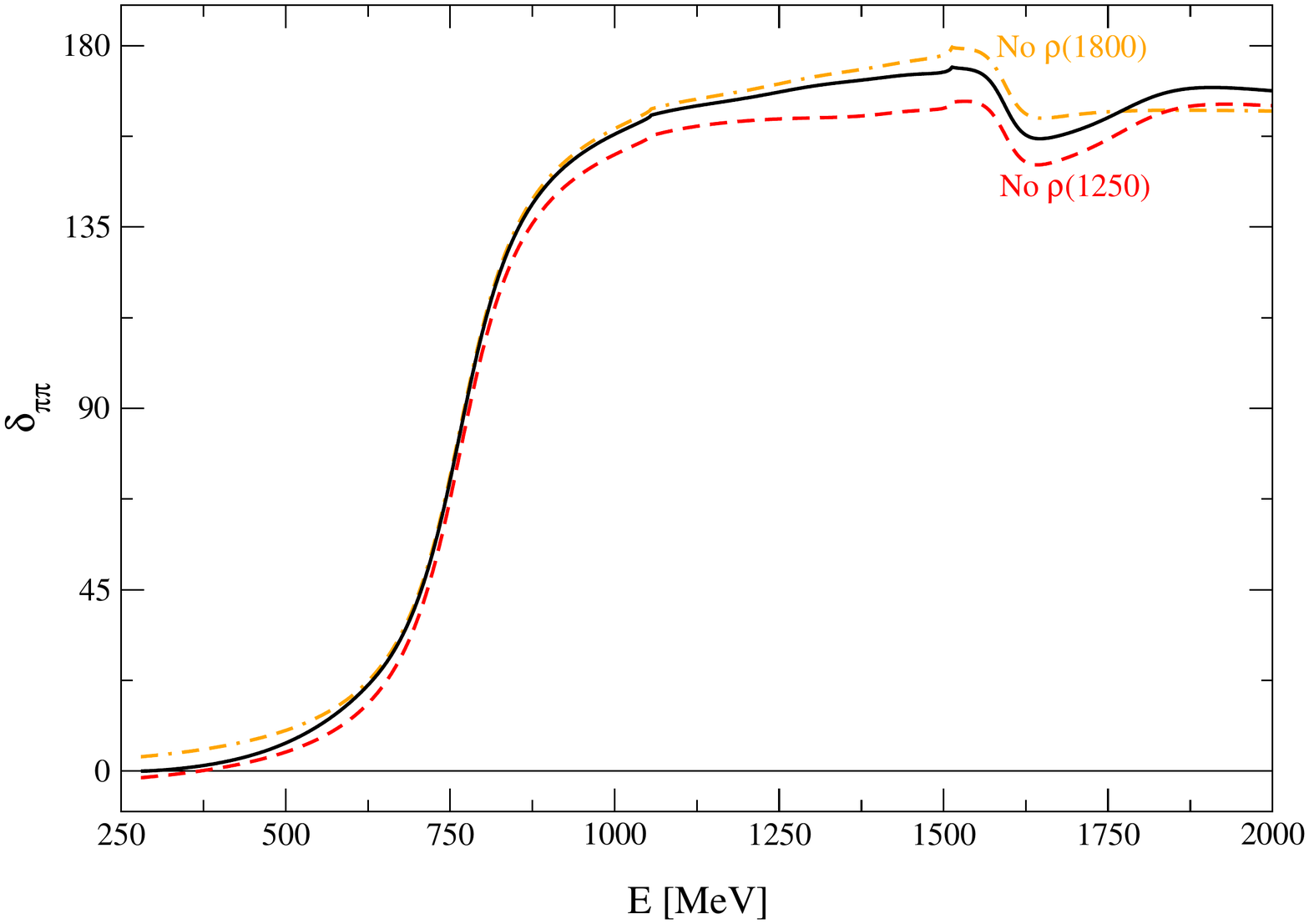}\\
\includegraphics[angle=0,scale=0.34]{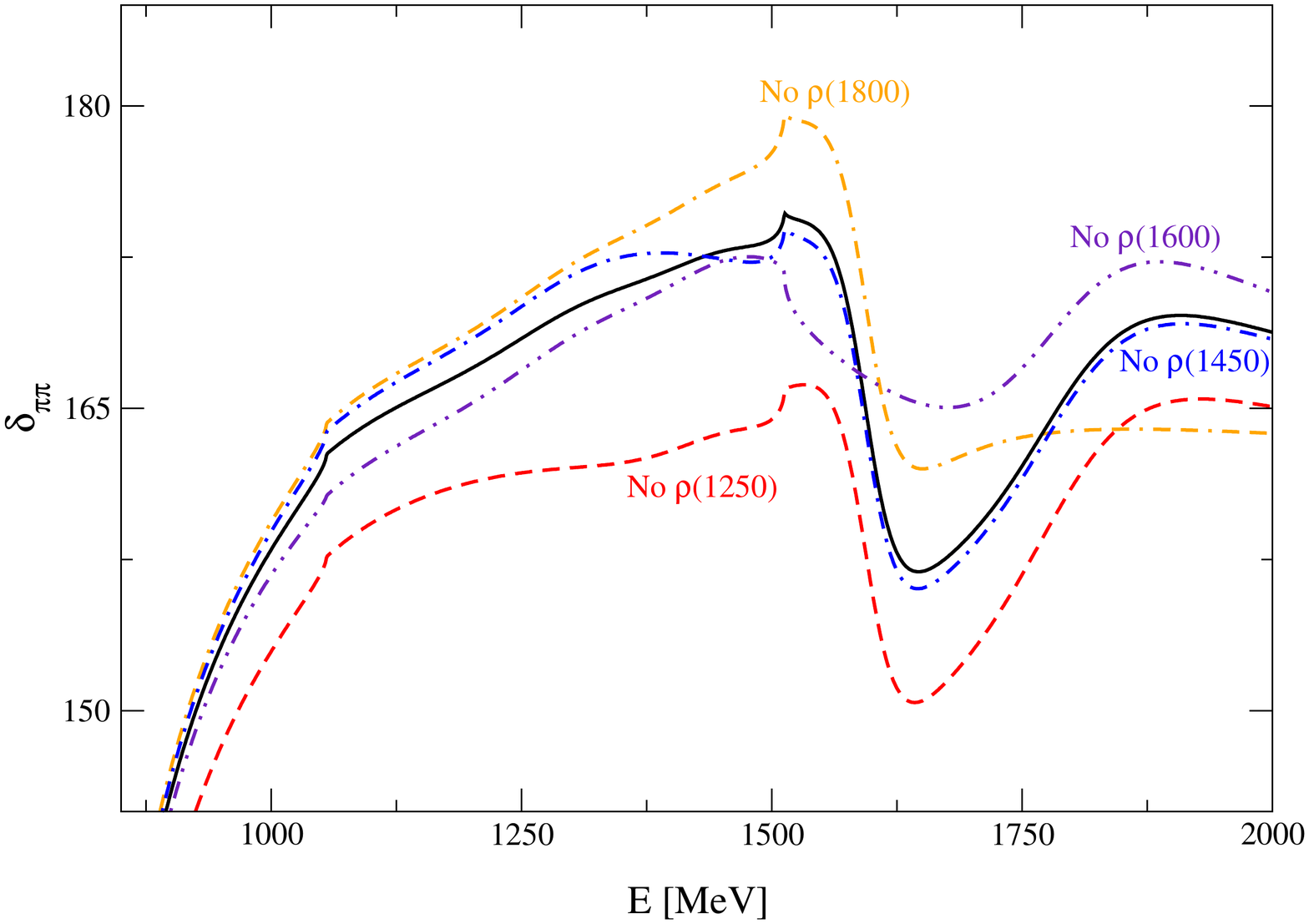}
\caption{Upper figure: phase shift from full amplitude (solid line)
and the same but without $\rho(1250)$ and $\rho(1800)$; lower figure:
as upper figure but enlarged over a reduced energy interval and also without
$\rho(1450)$ and $\rho(1600)$.}
\label{DeltaNoRho}
\end{center}
\end{figure}

\begin{figure}[!h]
\begin{center}
\includegraphics[angle=0,scale=0.34]{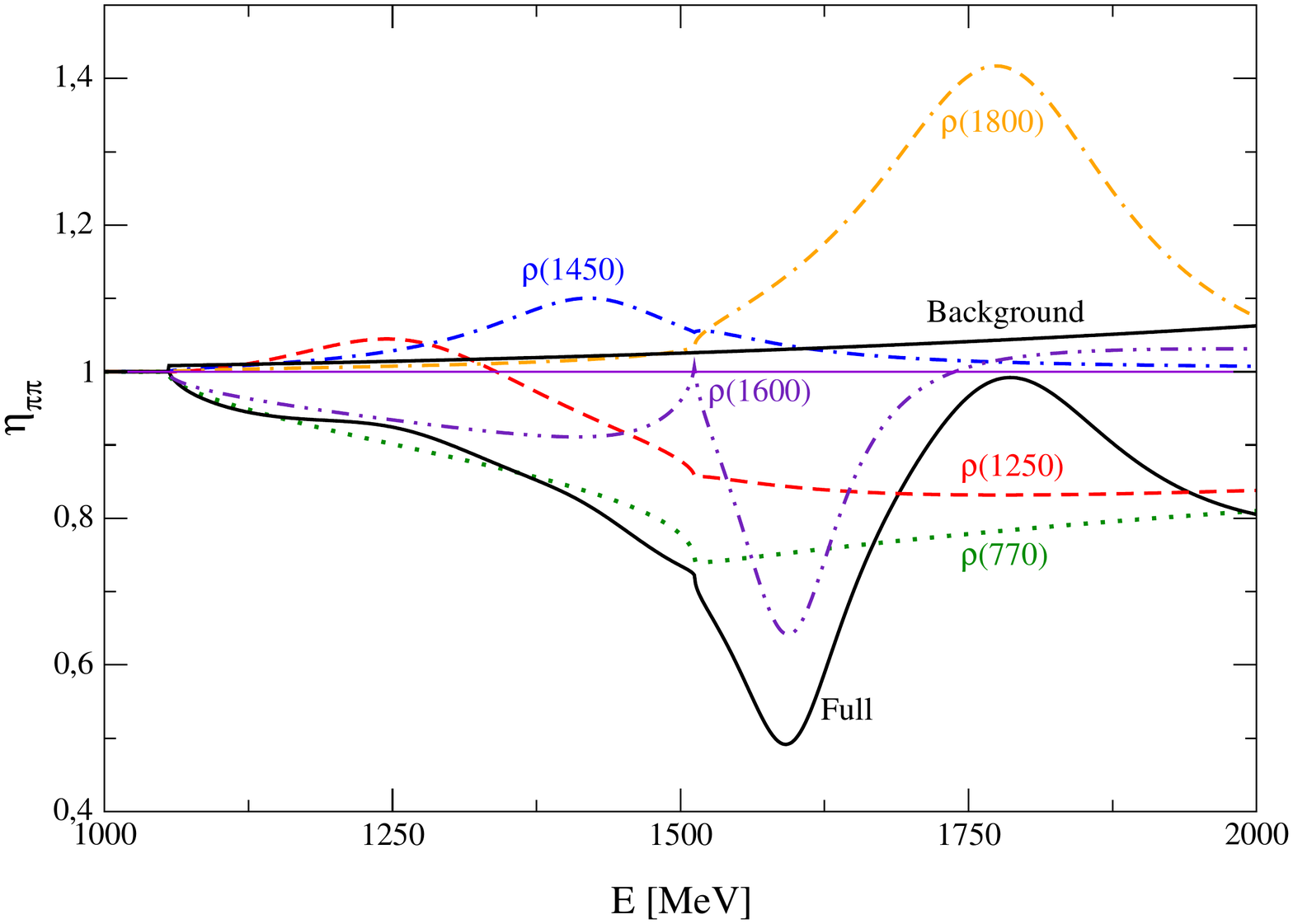}
\caption{Inelasticity due to individual resonances, full amplitude, and background.}
\label{ClasterEta}
\end{center}
\end{figure}

\begin{figure}[!h]
\begin{center}
\includegraphics[angle=0,scale=0.34]{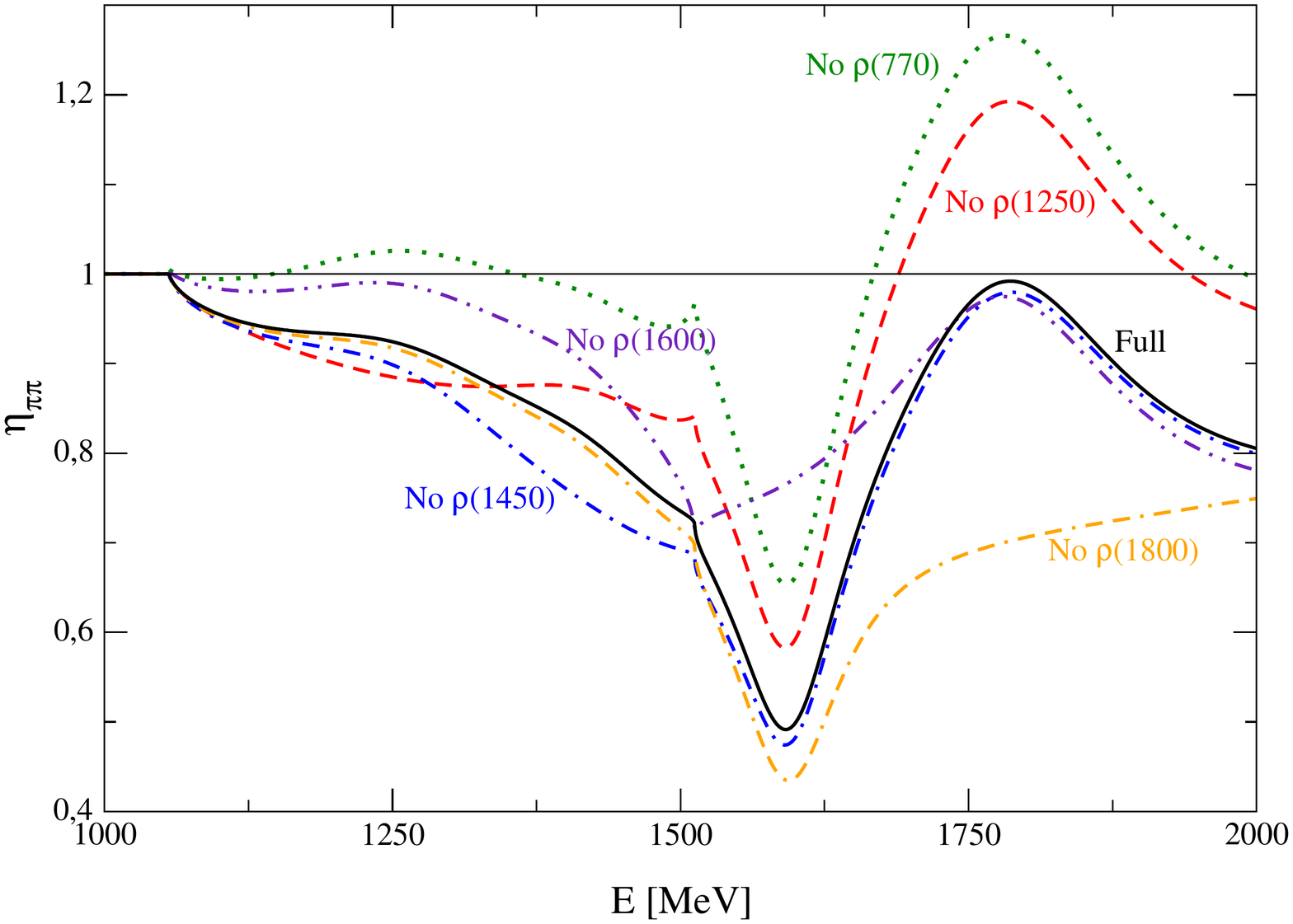}
\caption{Inelasticity due to full amplitude and same without individual
resonances.}
\label{EtaNoRho}
\end{center}
\end{figure}

The results presented in Figs.~\ref{FitDeltaAndEtaWithData}--\ref{EtaNoRho} are
based on the fits carried out in Refs.~\cite{PRD94p116013,Nazari-thesis}. 
Figure~\ref{FitDeltaAndEtaWithData} displays the $\pi\pi$ phase shifts and
inelasticities fitted to the also shown experimental data. Inspecting this
figure and Table~\ref{Tab_Chi2}, we clearly see that the curves reproduce the
data very well, all the way from the corresponding thresholds up to almost
2~GeV, especially when considering the small errors of the phase shifts.
Experimental data are from \cite{Batley2008}-\cite{Hyams75} for the phase shifts and from \cite{Batley2008}-\cite{Hyams73},\cite{Hyams75} for inelasticity.

Figure~\ref{ClastersDelta} shows the $P$-wave $\pi\pi$ phase shifts due to
the individual resonances, corresponding to poles (all members of a cluster
for a given resonance) on different Riemann sheets.
Of course, only the full phase shift has the correct threshold behavior,
given by a polynomial with fixed scattering length and effective range. 
As one would expect, $\rho(770)$ has by far the largest influence on the
overall phase, dominating the contributions of the $\rho$ excitations.
Moreover, the second most important resonance is clearly $\rho(1250)$, whereas
the smallest effect is due to $\rho(1450)$.
The visible yet rather insignificant kinks in the
phase-shift and inelasticity curves right above 1.5~GeV correspond to the
opening of the sharp effective $\rho\rho$ threshold. Clearly, they do not
affect the quality of the fits at all. Nevertheless, from a theoretical point
of view it would be desirable to somehow smear out this threshold so as to
account for the $\rho$ width, and the same for the $\rho2\pi$ channel.
An empirical way to do this was formulated in Ref.~\cite{EPJC71p1762}, by
allowing for complex masses in the final state. The resulting violation of
$S$-matrix unitarity was then corrected by redefining $S$ with the help of
a factorization valid for an arbitrary complex symmetric matrix. This may be
the subject of future work along the lines of the present analysis.

To properly assess the contribution of individual resonances to the full
amplitude, it is very clarifying to compute it before and after removing those
resonances. Figure~\ref{DeltaNoRho} displays phase shifts for the full
amplitude and that without terms from individual resonances. As the amplitude
of $\rho(770)$ is strongly dominant, especially below 1~GeV, changes by
removing this resonance are not shown, because they would be too large.
Once again one can clearly see how important the role of $\rho(1250)$ is, 
in contrast with most notably $\rho(1450)$. Its influence dominates between
1.0~GeV and 1.5~GeV, being comparable to that of $\rho(1600)$ and $\rho(1800)$
thereabove. The $\rho(1450)$ contribution is quite small over the entire
tested energy range.

Figure~\ref{ClasterEta} shows the $\pi\pi$ inelasticity $\eta$ for the
full amplitude and also the individual resonances. 
One can see very well that even below 1.5~GeV (near the $\rho\rho$ threshold)
inelasticity due to the $\rho(1250)$ amplitude significantly differs from 1
and together with part from that of $\rho(770)$ almost completely determines
the inelasticity of the full amplitude.
Contributions from $\rho(1450)$ and $\rho(1600)$ largely cancel each other
between the $\rho2\pi$ and $\rho\rho$ thresholds.
Above roughly 1.5~GeV, $\rho(1800)$ determines the energy dependence of
$\eta$ almost entirely, interfering with the still large but already rather
unstructured $\rho(1250)$ part comparable in size to that of $\rho(770)$.
The contribution of $\rho(1450)$ to $\eta$ is very small above 1.5~GeV. 
As expected, it only has a minor maximum at about 1.4~GeV.
The significant drop in the full inelasticity at about 1.6~GeV is mostly
determined by $\rho(1600)$, after the opening of the $\rho\rho$
channel. The role of the background in building $\eta$ is small, showing a
slow and smooth rise.

Just as in the case of the phase shift, the energy dependence of the
inelasticity for the amplitude without a given resonance, i.e., by omitting all
poles associated with it on the different Riemann sheets, is very informative.
In Fig.~\ref{EtaNoRho}, we see that removing $\rho(1250)$ would cause the
largest change  (after that caused by $\rho(770)$) to the inelasticity curve as compared to the one due to the full
amplitude. Similarly, a significant modification would be caused by leaving out
$\rho(1600)$ or $\rho(1800)$, but only around 1600~MeV or thereabove,
respectively. Finally, also here we observe that without $\rho(1450)$ there
would only be a modest change to $\eta$, over a relatively small energy region
below 1.5~GeV, having little effect on the shape of the inelasticity curve. 

Considering the relationship between resonances and the
inelasticity, one should refer to Eqs.~(\ref{Eq_Sfull})--(\ref{Eq_d}).
According to  unitarity, the Jost functions $d_{res}$ in Eq.~(\ref{Eq_d})
are constructed in such a way that their ratio in the unitary $S$-matrix
(Eqs.~(\ref{Eq_A},\ref{Eq_Sfull})) has a modulus equal to one in the whole
elastic region. This is due to the full symmetry between the poles and the
zeros there. In the inelastic region ($s> s_2$), this symmetry is
automatically removed ($w$ is no longer purely imaginary) and the moduli of
numerators and denominators are not equal anymore. The large number of
complex poles (characteristic of such a multiresonance analysis in a
three-channel approach) needed to describe the data and to meet the
crossing-symmetry condition naturally leads to various interferences among
all these poles, resulting in a total inelasticity consistent with unitarity
and the conditions imposed on the fits. Of course, the moduli of the
$d_{res}^*(-w^*)/d_{res}(w)$ ratios corresponding to individual resonances
(i.e., clusters of corresponding poles) do not have to fulfill the unitarity
condition of being smaller than 1, as can be seen in Fig.~\ref{ClasterEta}.
The same is true for the tiny contribution to the inelasticity due to the
background term defined in Eq.~(\ref{Eq_bgr}). Only mutual interferences 
among all resonances (poles) and the background produce the physical (unitary)
result. The sometimes reported unitary inelasticity of a particular resonance
corresponds to such a physical result (full $\eta$ in Fig.~\ref{ClasterEta}),
but is limited and calculated in the energy range selected for a given
resonance.

\section{\boldmath{$\rho(1250)$} from an analysis of the pion electromagnetic
form factor}
\label{em}
Vector isovector mesons below 2~GeV play a very important role in, for example,
the determination of the pion electromagnetic (EM) form factor.
When analyzing cross sections of $e^+ e^- \to \pi^+\pi^-$ production, this form
factor $F_{\pi}^{\smr{EM},I=1}(s)$ appears explicitly and contains information
on the dynamics of all these mesons:
\begin{eqnarray}
\sigma_{\smr{tot}}(e^+ e^- \to \pi^+\pi^-) & = &
\frac{\pi \alpha^2(0)}{3s} \beta_{\pi}^3(s) \times \nonumber \\
& & \hspace{-3cm}\left|F_{\pi}^{\smr{(EM)}I=1}(s)+Re^{i\phi}\frac{m_{\omega}^2}
{m_{\omega}^2-s-im_{\omega}\Gamma_{\omega}} \right |^2, 
\end{eqnarray}
where the pion ``velocity'' $\beta_{\pi}(s) = \sqrt{\frac{s-4m_{\pi}^2}{s}}$,
$R$  is the amplitude for $\rho$-$\omega$ mixing interference,
and the phase
$\phi=\arctan{\frac{m_{\rho}\Gamma_{\rho}}{m_{\rho}^2-m_{\omega}^2}}$
is the $P$-wave $\pi\pi$ phase shift determined at $s = m_{\omega}^2$.

An analysis in Ref.~\cite{PRD96p113004} compared two different approaches to
determining $F_{\pi}^{\smr{(EM)}I=1}(s)$. The first one was based on the
popular Gounaris-Sakurai (GS) \cite{PRL21p244} model constructed by
assuming that, for a wide energy range of the elastic region up to 1~GeV,
the $P$-wave isovector $\pi\pi$ scattering phase shift satisfies a
two-parameter effective-range formula of the Chew-Mandelstam type, i.e.,
\begin{equation}
\label{Eq_Chew_Mand}
\frac{q}{\sqrt{s}}\cot\delta = a+bq+q^2h(s) ,
\end{equation}
where $q$ is the pion momentum in the CM system and $h(s)$ is a simple
logarithmic function of $q$ and $s$. The pion EM form factor is then given by
\begin{equation}
\label{Eq_PionFF}
F_{\pi}^{\smr{(EM)}I=1}(s) = \frac{\sqrt{s}}{q^3}\frac{1}{\cot\delta(s)-i}.
\end{equation}
Using the fact that at $\delta = \pi/2$ 
the left-hand side of Eq.~(\ref{Eq_Chew_Mand}) vanishes and, comparing with
a Breit-Wigner distribution formula, the first derivative of the phase
shift can be given by $1/{m_{\rho}\Gamma_{\rho}}$, one can express
$F_{\pi}^{\smr{(EM)}I=1}(s)$ directly in terms of the $\rho(770)$  mass and
width.

The other approach employed in Ref.~\cite{PRD96p113004} was based on a very
simple unitary and analytic (UA) formula with two symmetric poles and zeroes
representing a resonance.
The left-hand cut was simulated with one pole (and a symmetric zero). 
In this way, it was shown \cite{PRD96p113004} that fits to
$e^+ e^- \to \pi^+\pi^-$ experimental cross sections and to elastic
$\pi\pi$ phase shifts (from GKPY equation) made independently using
the GS model and the UA approach give very different results
for the $\rho(770)$ resonance parameters, especially its mass.
As one can see in Table~\ref{Tab_mass_rho_elastic},
the mass difference is much larger than the estimated errors, which are
similar in size to those in the PDG tables \cite{PDG2020}.
\begin{table}[ht]
\begin{center}
\begin{tabular}{|c|c|c|}
\hline
 & GS & UA \\
\hline
$M_{\rho(770)}$ & $774.1 \pm 0.1$~MeV & $763 \pm 0.5$~MeV \\
$\Gamma_{\rho(770)}$ & $140.2 \pm 0.1$~MeV & $143.9 \pm 0.8$~MeV \\
\hline
\end{tabular}
\end{center}
\caption{Mass and width of $\rho(770)$ as calculated in the GS model
and the UA approach \cite{PRD96p113004} (see text).}
\label{Tab_mass_rho_elastic}
\end{table}
  
More significant differences were found in Ref.~\cite{PRD96p113004} for
the two higher $\rho$ states, in fits to cross sections up to pion momentum
squared $s=9$~GeV$^2$.
The authors of this analysis pointed out that a generalization of the GS
model above the inelastic threshold done in many experimental works
\begin{quote} \em
``is without any deeper physical background, as the original G.-S.\ model
for the $\rho^0$ meson contribution was constructed from the $P$-wave isoscalar
$\pi\pi$ scattering phase shift given by the generalized effective-
range formula of the Chew-Mandelstam type, which is
evidently valid only in the elastic region.'' \em
\end{quote} 
Nevertheless, to check the quantitative differences between the GS model and
the UA approach above 1 GeV, the authors did as in some experimental analyses,
i.e., they carried out fits to the data with three $\rho$ states, viz.\
$\rho(770)$, $\rho(1450)$, and $\rho(1700)$ using the GS model. Then they
compared the results to those obtained from their enhanced unitary and analytic
amplitude with poles as degrees of freedom. The results are presented in
Table~\ref{Tab_mass_3rhos} and compared to those from the PDG \cite{PDG2020}.
It is worth noting that the number of free parameters in the UA analysis is
smaller (11) than in the GS model (14). And despite the fact that the values
of $\chi^2$ per degree of freedom are 1.84 (UA) and 0.98 (GS), the results of
the UA analysis are much more realistic. In particular, we should draw
attention to the 171~MeV lower $\rho(1450)$ mass found in the UA approach as
compared to the GS model, which difference is much larger than the reported
errors in Table~\ref{Tab_mass_3rhos}.
\begin{table}[hb]
\begin{center}
\hspace{0.cm}\begin{tabular}{|c|c|c|c|}   
\hline
Parameter & PDG (MeV) & GS (MeV) & UA (MeV) \\
\hline
$m_{\rho(770)}$ & 775.26 $\pm$ 0.25 & 774.81 $\pm$ 0.01 & 763.88 $\pm$ 0.04\\
$m_{\rho(1450)}$ & 1465.00 $\pm$ 25.00 & 1497.70 $\pm$ 1.07 &  1326.35 $\pm$ 3.46\\
$m_{\rho(1700)}$ & 1720.00 $\pm$ 20.00 & 1848.40 $\pm$ 0.09  & 1770.54 $\pm$ 5.49\\
$\Gamma_{\rho(770)}$ & 147.80 $\pm$ 0.90 & 149.22 $\pm$ 0.01 & 144.28 $\pm$ 0.01\\
$\Gamma_{\rho(1450)}$ & 400.00 $\pm$ 60.00 & 442.15 $\pm$ 0.54 & 324.13 $\pm$ 12.01\\
$\Gamma_{\rho(1700)}$ & 250.00 $\pm$ 100.00 & 322.48 $\pm$ 0.69 & 268.98 $\pm$ 11.40\\
$\chi^2$ pdf & & 0.98 & 1.84 \\
& & 14 param. & 11 param. \\
\hline
\end{tabular}
\end{center}
\caption{Comparison of parameters for three $\rho$ states in the fits done
in Ref.~\cite{PRD96p113004} above the inelastic threshold up to pion momentum
squared $s=9$~GeV$^2$. For a better comparison, the values from the PDG tables
\cite{PDG2020} are also given.}
\label{Tab_mass_3rhos}
\end{table}

Similarly, significant differences can be found in the literature for the
mass of the very same resonance, which can often be easily explained 
by comparing their values found in e.g.\ BW and simple ``pole'' approaches. 
In a BW amplitude, the mass of a resonance is determined by the energy
$M_{\smr{BW}}$ at which the phase shift passes
90$^\circ$ ($M_{\smr{BW}} = 2\sqrt{k_{\smr{BW}}^2+m^2}$ for two interacting
equal-mass particles). The unitary ``pole'' amplitude (like the UA one used
above and in Ref~\cite{PRD96p113004}) has, for one single resonance, two
symmetric poles (and corresponding zeros) at $k_r = \pm a-ib$. Then, the
phase shift is given by
$\delta=\arctan{\frac{2bk}{a^2+b^2-k^2}}$
and it is clear that the value $\delta=90^\circ$ is attained for
$k \ne k_{\smr{BW}}$ (i.e., $2\sqrt{a^2+m^2} \ne M_{\smr{BW}}$). This
difference gets larger according as $b$ increases.
For $\rho(770)$ difference between $M_{BW}$ and "pole" mass $M_r$ defined by real part of $2\sqrt{k_r^2+m^2}$ is about 8~MeV, which explains the discrepancy seen in
Table \ref{Tab_mass_rho_elastic}
and found in many other theoretical and experimental analyses
\cite{PDG2020}. For $\rho(1450)$ this discrepancy is several dozen MeV,
which is less than the differences seen in
Table \ref{Tab_mass_3rhos}.
However, above 1~GeV one is dealing with a few very broad and highly
inelastic $\rho$ resonances and a simple reasoning like for $\rho(770)$ is 
completely insufficient, requiring to also account for other phenomena typical
of resonance interference.
It is worth noting here that by making fits using the BW and "pole" approaches and comparing $M_{BW}$ and $M_r$ for the $\rho(770)$ one can additionally request that the decay width be the same in both approaches. Then the difference between these masses is about 5 MeV.

A very good example of such an analysis, which is both qualitative and
quantitative in explaining the phase-shift behavior around 1250~MeV and the 
determination of the pole position in the amplitude, is Ref.~\cite{ZPC29p107}.
In this work entitled {\it ``Why is the $\rho'(1250)$ not Observed in the
$\pi\pi$ Scattering~?''}, an $M$-matrix parametrization of the $\pi\pi$ and
$\omega\pi$ partial amplitudes is analyzed and compared with the
parametrization as a sum of the inelastic resonance term and the background
amplitude. A most important qualitative conclusion is that even a small
inelastic background around 1300~MeV can completely hide a $\rho(1250)$ in
the $\pi\pi$ channel, leading to a non-resonant behavior there of the $\pi\pi$
phase shifts. The quantitative results of this analysis confirm this
conclusion: in order to describe the experimental data very well, a background
of only a few degrees around 1300~MeV is sufficient, the real part of the pole
then comes out at about 1220~MeV, the width is roughly 320~MeV, and 
$\Gamma_{\pi\pi}/\Gamma_{\omega\pi} \approx 0.15$, the latter ratio being in
reasonable agreement with the available experimental data \cite{PDG2020}.

\section{Conclusions}
\label{conclusions}
The results of our combined analyses unmistakable demonstrate the
necessity to include a $\rho^\prime$ resonance at about 1.26~GeV.
The stability of the fitted pole positions as well as the manifest
fulfillment of multichannel unitarity and optimized
crossing symmetry in our approach lend strong support to the reliability
of our excited $\rho$ states, including the ones at about 1.42~GeV,
1.60~GeV, and 1.78~GeV. Straightforward spectroscopic arguments then
impose the following quark-model assignments:
$\rho(1250)$/$2\,^{3\!}S_1$,
$\rho(1450)$/$1\,^{3\!}D_1$,
$\rho(1600)$/$3\,^{3\!}S_1$, and
$\rho(1800)$/$2\,^{3\!}D_1.$
Confirmation of these four states, which were already
found in a previous analysis \cite{PRD81p016001}, poses serious
problems to mainstream quark models, unless at least $\rho(1250)$ is
interpreted as a crypto-exotic tetraquark state, for which there is
no experimental or theoretical support  (also see the discussion of
``$\omega_x$'' above). A $\rho^\prime$ at 1.25~GeV is very hard to
reconcile with the GI \cite{PRD32p189} model and similar ones, based on a
Coulomb-plus-linear confining potential with a running strong coupling
constant $\alpha_s(q^2)$. The only way out would be to consider the GI
$\rho(1450)$ a ``bare'' quark-antiquark bound state that upon unitarization
turns into the physical $\rho(1250)$ resonance, similarly as in the
unitarized quark model of Ref.~\cite{PRD27p1527}. However, the latter model
also predicts mass shifts for all other states, especially the ground state
$\rho(770)$, besides employing a completely different confinement mechanism.
Therefore, the complete spectrum including unitarization effects would have
to be computed again in the GI and similar models, after refitting the
parameters. But even more seriously, it is practically inconceivable that the
$3\,^{3\!}S_1$ state in the GI model at 2.0~GeV could be lowered by
unitarization to 1.6~GeV. The problem is that the radial splitting between the
first and second excitation for mesons with light quarks in the GI model is
larger than 500~MeV \cite{PRD32p189}. This gives rise to huge discrepancies
with the observed spectra \cite{PDG2020} not only for vector states, but also
for tensors like the $f_2$ \cite{APPBPS5p1007,PRD81p016001}, as already
mentioned in Sec.~\ref{intro}. Therefore, one must either present very
convincing arguments why some of the resonances identified in the present and
previous \cite{NPA807p145,PRD81p016001} analyses should be interpeted as
crypto-exotic states, or consider the possibility that the Coulomb-plus-linear
confining potential with a running coupling constant is inadequate, at least
in the way it is usually implemented in quark models.

Regardless of these considerations, we believe to have made a convincing
case for $\rho(1250)$, which should finally be rehabilitated in the PDG
tables with a separate entry. Further expertimental analyses that do
respect multichannel $S$-matrix unitarity would be most welcome, of course,
besides realistic model and lattice calculations accounting for unitarization
effects.

The problem with new experimental data is that they will
most likely result from production processes and not elastic
scattering, making their direct inclusion in an analysis
as described in the present paper problematic. A possible
way out would be applying the formalism for relating production
and scattering amplitudes as outlined in Ref.~\cite{AP323p1215}
to our multichannel $S$-matrix approach. Also this will be a topic
of future research.

\section*{Acknowledgments}
We are indebted to E.~van~Beveren for providing us with many references
concerning the $\rho(1250)$ resonance and one of us (G.R.) also for
prior collaboration and many discussions on this topic.
This work was partly financed by a Polish research project with
no.~2018/29/B/ST2/02576 (National Science Center).

\end{document}